\documentclass[a4paper, amsfonts, amssymb, amsmath, reprint, showkeys, nofootinbib, twoside]{revtex4-1}
\usepackage[english]{babel}
\usepackage[utf8]{inputenc}
\usepackage[colorinlistoftodos, color=green!40, prependcaption]{todonotes}

\usepackage[pdftex, pdftitle={Article}, pdfauthor={Author}]{hyperref} 
\usepackage{hyperref}
\usepackage{verbatim}
\usepackage{amsmath}
\usepackage{amsfonts}
\usepackage{amssymb}
\usepackage{amsthm}
\usepackage{bm}
\usepackage{graphicx}
\usepackage{xcolor}
\usepackage{textcase}
\usepackage{url}
\usepackage{appendix}
\usepackage{epstopdf}

\def\av#1{\langle#1\rangle}

\newcommand{\diagdots}[3][-25]{%
  \rotatebox{#1}{\makebox[0pt]{\makebox[#2]{\xleaders\hbox{$\cdot$\hskip#3}\hfill\kern0pt}}}%
}

\begin{document}
\title{Bell's theorem for trajectories}

\author{Dragoljub Go\v{c}anin}
  \email[Correspondence email address: ]{dgocanin@ipb.ac.rs}
    \affiliation{Faculty of Physics, University of Belgrade, Studentski Trg 12-16, 11000 Belgrade, Serbia}
\author{Aleksandra Dimi\'{c}}
  \affiliation{Faculty of Physics, University of Belgrade, Studentski Trg 12-16, 11000 Belgrade, Serbia}
\author{Flavio Del Santo}
\affiliation{Vienna Center for Quantum Science and Technology (VCQ), Faculty of Physics,
Boltzmanngasse 5, University of Vienna, Vienna A-1090, Austria and
Institute for Quantum Optics and Quantum Information (IQOQI),
Austrian Academy of Sciences, Boltzmanngasse 3, A-1090 Vienna, Austria}
\author{Borivoje Daki\'{c}}
\affiliation{Vienna Center for Quantum Science and Technology (VCQ), Faculty of Physics,
Boltzmanngasse 5, University of Vienna, Vienna A-1090, Austria and
Institute for Quantum Optics and Quantum Information (IQOQI),
Austrian Academy of Sciences, Boltzmanngasse 3, A-1090 Vienna, Austria}

\date{\today}

\begin{abstract}

In classical theory, the trajectory of a particle is entirely predetermined by the complete set of initial conditions via dynamical laws. Based on this, we formulate a no-go theorem for the dynamics of classical particles, i.e., {\it a Bell's inequality for trajectories}, and discuss its possible violation in a quantum scenario. A trajectory, however, is not an outcome of a quantum measurement, in the sense that there is no observable associated with it, and thus there is no ``direct'' experimental test of the Bell's inequality for trajectories. Nevertheless, we show how to overcome this problem by considering a special case of our generic inequality that can be experimentally tested point-by-point in time. Such inequality is indeed violated by quantum mechanics, and the violation persists during an entire interval of time and not just at a particular singular instant. We interpret the violation to imply
that trajectories (or at least pieces thereof) cannot exist predetermined, within a local-realistic theory.
\end{abstract}
\maketitle

\section{Introduction}

The essential feature of classical mechanics is that successive positions of a point-like particle constitute a continuous trajectory that is uniquely defined by dynamical equations 
together with the appropriate set of initial conditions.
Bell's theorem \cite{Bell64}, on the other hand, demonstrates that quantum theory is incompatible with the outcomes of measurements being predetermined. In fact, the violation of Bell's inequalities guarantees that there cannot exist any local-realistic 
model that accounts for the observed statistics. A typical Bell's scenario features two distant (spacelike separated) observers, Alice and Bob, who each performs local measurements on their respective system (these two systems may have interacted in the past). It is assumed that each of them freely and independently 
picks measurement settings (or inputs), $a$ and $b$ for Alice and Bob, and obtain outcomes $\alpha$ and $\beta$, respectively. The assumption of ``local realism'' means that the probability distribution of the respective local outcomes are conditionally independent (i.e., locally factorable), given that one takes into account all the possible ``hidden variables'' $\lambda$, i.e., 
\begin{equation}\label{bellfact}
p(\alpha,\beta|a,b)=\int_\Lambda d\lambda\; \mu(\lambda) p(\alpha|a,\lambda)p(\beta|b,\lambda).
\end{equation}
The mutual dependence between measurement outcomes is solely due to the lack of experimental control (lack of knowledge) of the full set of parameters $\lambda\in\Lambda$, which are distributed according to some probability distribution $\mu(\lambda)$. This form of factorization can be used to derive no-go theorems in the form of inequalities (known as \emph{Bell's inequalities}) which put strict bounds on possible statistics of measurable quantities. As such, within a theory that predicts a violation of Bell's inequalities, local realism cannot be upheld. Quantum mechanics is indeed an example of such a theory and it is by now a corroborated experimental result that Bell's inequalities can be violated by using entangled quantum states \cite{Christensen, Hensen, Shalm}.

As a concrete example, consider the simplest Bell's inequality (known as the Clauser-Horne-Shimony-Holt inequality, after the physicists who put it forward \cite{CHSH}), where both inputs and outputs are binary variables, i.e., $a, b\in \{0,1\}$ and $\alpha, \beta\in\{-1,1\}$. Then the condition of local realism \eqref{bellfact} implies the inequality
\begin{equation}\label{belldef}
S:=\langle \alpha \beta\rangle_{00}+\langle \alpha \beta\rangle_{01}+\langle \alpha \beta\rangle_{10}-\langle \alpha \beta\rangle_{11} \leq 2,
\end{equation}
where $\langle \alpha \beta\rangle_{ab} = \sum_{ \alpha, \beta}  p(\alpha,\beta|a,b)\alpha \beta$ are the correlations between measurement outcomes, given the inputs. Quantum mechanics allows us to violate this inequality, reaching a maximal value of $S=2\sqrt2$ (known as \emph{Tsirelson's bound} \cite{Tsirelson}).

In the same spirit, the aim of this paper is to derive a (testable) no-go theorem that rules out trajectories of particles in quantum mechanics. One may object that trajectories are a purely classical concept in the first place, and simply cannot be translated into quantum mechanics due to the uncertainty principle, which sets a fundamental limit to the joint determinacy of position and momentum. However, this can be regarded as reflecting a lack of knowledge about the underlying \emph{true state of affairs}. Thus, the conclusion that the uncertainty principle alone implies that particles' trajectories cannot exist in the quantum regime could be upheld only if quantum mechanics is a complete theory. But what if quantum mechanics could be completed by some additional \emph{hidden variables} that would overcome the uncertainty principle and retrieve the concept of trajectory?
In Ref. \cite{HVmodel}, Bell constructed a concrete hidden variable model for a single qubit that predicts the same statistics of spin measurement outcomes as quantum mechanics. And the acclaimed Bohm's hidden variable theory \cite{Bohm1, Bohm2}-- despite being admittedly nonlocal--while giving the same predictions as quantum mechanics, allows us to speak realistically of particles' trajectories.

It ought to be stressed, however, that the assumption of the existence of a predetermined trajectory of a particle alone cannot be ruled out by quantum theory. As a matter of fact, in Sec. I of our Supplemental Material we provide a general argument for the possibility of a hidden variable
description of single-particle dynamics in terms of trajectories, whose predictions are in agreement with those
of quantum mechanics. Therefore, the ``assumption of trajectories'' can be tested against 
quantum-mechanical predictions only when supplemented with additional assumptions. One can, for example, consider ``macroscopic realism'' where the additional assumption is the noninvasive measurability and test the model via Leggett-Garg inequalities (LGIs) \cite{LG,LG1,LG2,LG3,LG4}. 
However, in this paper we appeal to Bell's inequalities
for disproving particle trajectories via Bell's ``local realism'', as defined in Eq. \eqref{bellfact}.

Although the formulation of a Bell's theorem for trajectories is relatively simple, its empirical confirmation is more challenging. The main problem is that, in the quantum theory, there is no observable associated with the trajectory of a particle. Hence, there is no straightforward means on how to directly measure it. Similar problems have been studied in the context of consistent histories \cite{griffiths, dowker} and, more specifically, entangled histories \cite{cotler1, cotler2}. It seems that the only experimentally accessible object is a single point of the trajectory, obtained by measuring the particle's position at a given instant of time. From this perspective, the trajectory can be seen as a sequence of such points during an interval of time. This observation will serve us to derive a whole class of experimentally accessible Bell's inequalities for trajectories, where the actual test is done in a point-by-point fashion. This ``local in time'' violation is indeed obtained in a quantum-mechanical setting. Moreover, the violation can hold continuously during an entire interval of time and not just at a particular instant, and it can thus be seen as the evidence that there are nonlocal quantum correlations at the level of the whole trajectories (or at least pieces thereof), thus disproving their local-realistic description. Our example involves two entangled particles whose dynamics are governed by local Hamiltonians. In this case, the local settings are, for each party, encoded in choices of the local potentials in the Hamiltonians. 

\section{Bell's inequality for trajectories}

\begin{figure}[h]
    \centering
   \includegraphics[width=8.2cm]{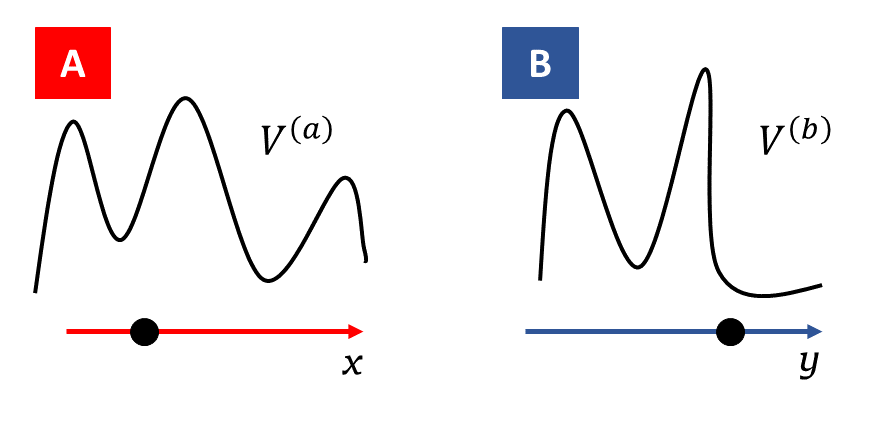}
\caption{Setting up local potentials: Alice ($A$) and Bob ($B$) locally encode their freely chosen binary inputs $(a,b)$ in potentials $V^{(a)}(x)$ and $V^{(b)}(y)$ that will govern the dynamics of their particles. Each particle is constrained to move along a line, parametrized by the $x$ coordinate for Alice's particle and the $y$ coordinate for Bob's.}
\label{potentials} 
\end{figure}

To derive Bell's inequality for trajectories, we consider the standard Bell's scenario. Two parties, Alice (A) and Bob (B), reside in two distant laboratories and each has a particle that they can manipulate locally. As is customary in Bell’s scenarios, also here the local measurements are performed in spacelike separation to prevent the possibility of local interactions. We assume that Alice and Bob constitute a pair of inertial reference frames that are at rest with respect to each other. Therefore, they have the same time coordinate (i.e., they share a clock).
Alice and Bob can freely and independently choose binary
settings (inputs), respectively labeled $a$ and $b$. They then locally encode their choices by specifying a set of dynamical parameters to govern the dynamics of their respective particles.
For example, A and B can encode their choices into potentials, $V^{(a)}$ and $V^{(b)}$, as shown in Fig. \ref{potentials}). For simplicity, we derive our Bell's inequality for one-dimensional (1D) trajectories; however, a generalization to higher dimensions is straightforward. 

Let us start by describing the evolution of one particle only, say, Alice's. As already recalled, in classical physics the trajectory of a particle is entirely defined by the complete set of initial conditions and dynamical laws. Suppose now that Alice, in her laboratory, has control over a certain set of parameters $a$ (which play the role of measurement settings in the standard Bell's experiments), and conducts an experiment to determine the trajectory of the particle. In a realistic scenario, however, Alice will in general lack control over some other relevant ``hidden'' parameters $\lambda_{A}$, e.g., controllable parameters $a$ could specify the Hamiltonian of the system and $\lambda_{A}$ could refer to the uncontrollable initial conditions $(x_0, p_0)$. Yet, it is necessary to specify the full set of parameters $(a,\lambda_{A})$ in order to deterministically characterize 
the unique trajectory $X^{(a)}_{\lambda_{A}}$ of the particle. 
Therefore, the probability (density) to get a particular trajectory $X$ in the time interval $[0,\tau]$ given the setting $a$ reads $p_{A}[X|a]=\int d\lambda_{A}\;\mu_{A}(\lambda_{A})\prod_{t=0}^{\tau}\delta[X(t)-X^{(a)}_{\lambda_{A}}(t)]$, with some probability distribution $\mu_{A}$ over all possible values of the hidden variables $\lambda_{A}$; and we have a similar expression in Bob's case.

Coming back to a bipartite scenario, one can, in a similar fashion, construct the joint probability distribution of the two trajectories, given the inputs ${a,b}$, by averaging over the hidden parameters $\lambda_A$ and $\lambda_B$ for $A$ and $B$ respectively. If the evolutions of the two particles are to be governed by their respective local Hamiltonians only (assumption of local realism), the joint conditional distribution is of the form
\begin{align}\label{local dist}
p[X,Y|&a,b]=
\int d\lambda_Ad\lambda_B\;\mu(\lambda_A,\lambda_B) \nonumber\\ &\times\prod_{t=0}^{\tau}\delta[X(t)-X^{(a)}_{\lambda_A}(t)]\delta[Y(t)-Y^{(b)}_{\lambda_B}(t)],
\end{align}
where $\mu(\lambda_A,\lambda_B)$ is the joint distribution of the hidden parameters. We now introduce the operation of averaging over the distribution of trajectories and consider functionals that take trajectories as inputs. For some functional of the difference of two trajectories, $F[X-Y]$, its mean value is given by \begin{align}\label{averages}
\av{F}_{ab}&=\int DX DY\; p[X,Y|a,b] F[X-Y] \nonumber\\ 
&= \int d\lambda_Ad\lambda_B\;\mu(\lambda_A,\lambda_B)F[X^{(a)}_{\lambda_A}-Y^{(b)}_{\lambda_B}],
\end{align}
which directly follows from local form of the conditional probability distribution provided in \eqref{local dist}. 

Consider now any symmetric (i.e. $F[X]=F[-X]$) and subadditive (i.e., $F[X+Y]\leq F[X]+F[Y]$) functional. One can write the following general Bell's inequality\footnote{We can generalize this inequality to include arbitrary subadditive functionals, not necessarily symmetric ones; in this case it reads  
$S=\sum_{a,b=0}^1(-1)^{ab}\av{F[(-1)^{(1-a)(1-b)}(X-Y)]}_{ab}\geq0$.}
\begin{equation}\label{Bell}
S:=\sum_{a,b=0}^1(-1)^{ab}\av{F}_{ab}\geq0,
\end{equation}
which follows directly from the triangle inequality (see Sec. II of our Supplemental Material). 
However, to evaluate the averages entering Eq. \eqref{Bell}, we would need the entire trajectories $X$ and $Y$ as outcomes of measurements, which is problematic in quantum theory, because trajectories are not observables (in a strict mathematical sense). This is an obstacle, even in principle, to directly test our Bell's inequality. However, it is reasonable to expect violation in a quantum setting, at least in some form. 

To make our Bell's inequality testable, we provide an operational meaning to these trajectory measurements in the following sense: Suppose \linebreak[4]
 $F[X-Y]=\int_{0}^{\tau}dtf[X(t)-Y(t)]$, where $f$ is a symmetric and subadditive function, i.e., $f(x-y)=f(y-x)$ and $f(x+y)\leq f(x)+f(y)$ (such as the norm distance $f(x-y)=|x-y|$). Clearly, this property induces the subadditivity of $F$, and thus Eq. \eqref{Bell} holds. The expression for averages in Eq. \eqref{Bell} now reads 
\begin{equation}\label{point av}
 \av{F}_{ab}=\int_{0}^{\tau}dt\av{f(t)}_{ab},
\end{equation} 
with $\av{f(t)}_{ab}=\int d\lambda_A d\lambda_B\;\mu(\lambda_A,\lambda_B)f[X^{(a)}_{\lambda_A}(t)-Y^{(b)}_{\lambda_B}(t)]$. Finally, the Bell's inequality \eqref{Bell} becomes
\begin{equation}\label{point Bell}
  S=\int_{0}^{\tau} dt\mathcal{S}(t)\geq0,
\end{equation}
where we introduced the time dependent \emph{Bell's parameter} $\mathcal{S}(t):=\sum_{a,b=0}^{1}(-1)^{ab}\av{f(t)}_{ab}$, satisfying the ``local in time'' inequality $\mathcal{S}(t)\geq 0$ for every $t$ (assuming local realism). This is actually a continuous family of Bell-like inequalities for the coordinates $x$ and $y$ of the pair of particles for each particular instant of time. The quantity $S$ is now experimentally testable, for it can be evaluated from point-by-point measurements of the Bell's parameter $\mathcal{S}(t)$ in time. We assume that Alice and Bob have synchronized clocks, as in the standard Bell's experiment, and they sample the function $\mathcal{S}(t)$ by performing position measurements on their respective particles at a common instant of time $t$ at each run of the experiment. In this view, the expression \eqref{point Bell} is understood as \emph{a Bell's inequality for trajectories}, because the violation of the inequality $\mathcal{S}(t)\geq 0$ for some finite continuous interval of time $[t_{i},t_{f}]$ during the evolution of the particles (not necessarily during the whole interval $[0,\tau]$) would rule out the possibility for the particles to have predetermined trajectories (more precisely, the pieces thereof that correspond to the interval $[t_{i},t_{f}]$ during which $\mathcal{S}(t)<0$), within any local-realistic theory. In fact, if $\mathcal{S}(t)<0$ during some interval $[t_{i},t_{f}]\subset[0,\tau]$, then we necessarily have $S<0$ at least for \emph{that} interval of time, and the corresponding pieces of trajectories cannot be accounted for by any local-realistic theory.

In what follows, we demonstrate, by using a simple dynamical model, that quantum mechanics can indeed allow this kind of violation.


\section{Quantum scenario}

Let us suppose that Alice and Bob share a pair of quantum particles, both of mass $M$, prepared in some pure initial state $\vert\Psi_{0}\rangle$. Alice and Bob then encode their freely-chosen inputs $a$ and $b$ in the potentials that will govern the dynamics of their particles. For a given pair of inputs $(a,b)$, we have a pair of Hamiltonians   
\begin{align}\label{Hams}
\hat{H}_{A}^{(a)}=\frac{\hat{p}_{x}^{2}}{2M}+\hat{V}^{(a)}(\hat{x}),\;\;\; \hat{H}_{B}^{(b)}=\frac{\hat{p}_{y}^{2}}{2M}+\hat{V}^{(b)}(\hat{y}).
\end{align}
Since there is no interaction between the particles during the evolution, their initial state evolves, in the Schr\"{o}dinger picture, as $\vert\Psi^{(a,b)}(t)\rangle=\hat{U}_{A}^{(a)}(t)\otimes\hat{U}_{B}^{(b)}(t)\vert\Psi_{0}\rangle$, \linebreak[4]
where $\hat{U}_{A/B}^{(a/b)}(t)=\exp\left(-{\frac{i}{\hslash}\hat{H}_{A/B}^{(a/b)}} t \right)$.

The ``quantum'' Bell's parameter thus reads
\begin{align}\label{BQM}
\mathcal{S}_{QM}(t)=\sum\limits_{a,b=0}^{1}(-1)^{ab}\langle \Psi^{(a,b)}(t)\vert \hat{f}(\hat{x}-\hat{y})\vert\Psi^{(a,b)}(t)\rangle.
\end{align}

Alternatively, and more appropriately for our purpose, we can switch to the Heisenberg picture. Therefore, we consider the time evolution of $\hat{f}(\hat{x}-\hat{y})$ for a given pair of inputs $(a,b)$, 
\begin{equation}\label{bpt}
\hat{f}_{ab}(t)=\hat{U}^{(b)\dagger}_{B}(t)\hat{U}^{(a)\dagger}_{A}(t)\hat{f}(\hat{x}-\hat{y})\hat{U}^{(a)}_{A}(t)\hat{U}^{(b)}_{B}(t).
\end{equation}
In this picture, the Bell's parameter reads
\begin{equation}
\label{BellQM}
\mathcal{S}_{QM}(t)=\langle\Psi_{0}\vert\hat{\mathcal{S}}_{QM}(t)\vert\Psi_{0}\rangle=\sum\limits_{a,b=0}^{1}(-1)^{ab}\langle\Psi_{0}\vert\hat{f}_{ab}(t)\vert\Psi_{0}\rangle.
\end{equation}

If for some particular instant of time $t=T$ there exists an eigenvalue of the operator $\hat{\mathcal{S}}_{QM}(T)$, as defined in \eqref{BellQM}, that is smaller than zero, then we can choose the corresponding eigenfunction to be the initial state $\vert \Psi_{0}\rangle$ and thus assure that $\mathcal{S}_{QM}(T)<0$. From the continuity of the Bell's parameter as a function of time, we expect that the local-realistic inequality $\mathcal{S}(t)\geq 0$ must also be violated by $\mathcal{S}_{QM}(t)$ in some neighborhood of $t=T$, i.e., for some finite, continuous interval of time around $T$.

The problem is thus reduced to finding an appropriate initial state $\vert\Psi_{0}\rangle$ together with the potentials $\hat{V}^{(a)}$ and $\hat{V}^{(b)}$ such that $\mathcal{S}_{QM}(t)<0$ during some finite, continuous interval of time, thus leading to the violation of Eq. (\ref{point Bell}), at least during that particular interval. 
As in standard Bell's inequalities, entanglement plays a crucial role in a violation of the Bell's inequalities for trajectories. Note, however, that since there is no observable associated with a full trajectory, entanglement comes into play through the choice of the initial state $\vert\Psi_{0}\rangle$. As a matter of fact, if a pair of particles is initially prepared in a separable state, the quantum Bell operator $\mathcal{S}_{QM}(t)$, defined in Eq. (\ref{BellQM}), is positive for every value of $t$. Thus, an initial separable state cannot be used to bring about the violation (see Sec. III of our Supplemental Material).  

\begin{figure}[h]
   \centering
   \includegraphics[width=8.4cm]{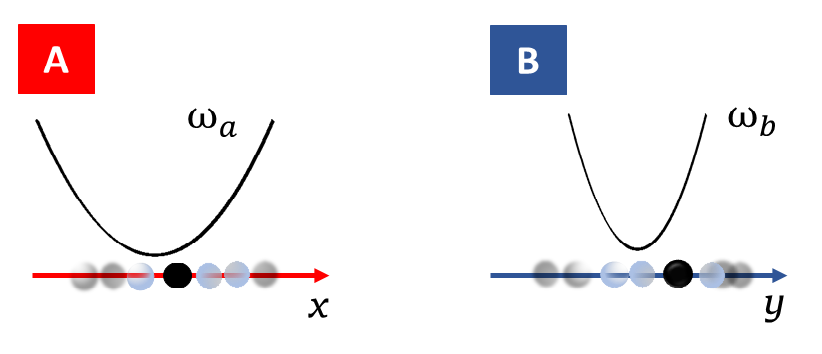}
   \caption{Setting up harmonic potentials: illustration of the Bell's experiment with a pair of quantum particles confined in harmonic potentials, regulated by the two parties Alice (A) and Bob (B). The binary inputs $(a,b)$ are encoded in the frequency parameters $(\omega_{a}, \omega_{b})$ of the potentials.}
\label{parabole} 
\end{figure}

\section{Example of violation}
To set up a concrete dynamical model that allows a violation of the Bell's inequality for trajectories, we use a pair of quantum harmonic oscillators. Alice and Bob locally encode their inputs $(a,b)$ by setting up harmonic potentials $\hat{V}_{A}^{(a)}=\tfrac{M\omega_{a}^{2}}{2}\hat{x}^{2}$ and $\hat{V}_{B}^{(b)}=\tfrac{M\omega_{b}^{2}}{2}\hat{y}^{2}$. More precisely, the inputs are encoded by tuning the frequency parameters $(\omega_{a}, \omega_{b})$ of the potentials (see Fig. \ref{parabole}).

As an ansatz for the initial state $\vert\Psi_{0}\rangle$ we consider a general state of two quantum harmonic oscillators, both having frequency $\Omega$, that belongs to the subspace spanned by the basis $\{\vert m\rangle_{A}\vert n\rangle_{B}\;\vert\; m,n=0,1,\dots,8\}$, i.e.,  
\begin{equation}
    \vert\Psi_{0}\rangle=\sum\limits_{m,n=0}^{8}c_{mn}\vert m\rangle_{A}\vert n\rangle_{B},
\end{equation}
with some \textit{a priori} undetermined amplitudes $c_{mn}$. We could, of course, take a more general ansatz, but in order to see the violation it turns out to be enough to consider only the first nine harmonics for each oscillator. Note that $\vert m\rangle_{A}$ and $\vert n\rangle_{B}$ are energy eigenstates of the respective oscillators for the frequency $\Omega$; they need not be eigenstates for the evolution operators $\hat{U}_{A}^{(a)}$ and $\hat{U}_{B}^{(b)}$ because these operators depend on the choice of the parameters $\omega_{a}$ and $\omega_{b}$. The relevant parameters of the system, $M$ and $\Omega$, provide the units of time and length. The unit of time is simply $\Omega^{-1}$, and the unit of length is $(\hbar/M\Omega)^{1/2}$. Together, they set the scale of the problem. For example, if we take the mass of an electron $M\sim 10^{-30}\;\mathrm{kg}$ and frequency $\Omega\sim 10^{8}\;\mathrm{rad/s}$, the relevant length scale is $(\hbar/M\Omega)^{1/2}\sim 1\;\mathrm{\mu m}$. 

In order to find a simple case of the violation of the Bell's inequality for trajectories, we consider here, among all symmetric and subadditive functions, the absolute value (Euclidean distance) $d(x-y)=\vert x-y\vert$. In coordinate representation, the corresponding operator satisfies $\langle x,y\vert\hat{d}(\hat{x}-\hat{y})\vert\Psi\rangle=\vert x-y\vert\Psi(x,y)$.
In this particular case, the time-dependent Bell's parameter reads
\begin{equation}
\mathcal{S}_{QM}(t)=\sum\limits_{a,b=0}^{1}(-1)^{ab}\langle\Psi_{0}\vert\hat{d}_{ab}(t)\vert\Psi_{0}\rangle,
\end{equation}
where $\hat{d}_{ab}(t)$ represents the specific choice of the operator $\hat{f}_{ab}(t)$ defined in Eq. \eqref{bpt}. The matrix elements of the Bell's operator $\hat{\mathcal{S}}_{QM}(t)$, generally defined in Eq. \eqref{BellQM}, in the reduced basis $\{\vert m\rangle_{A}\vert n\rangle_{B}\;\vert\; m,n=0,1,\dots,8\}$ are
\begin{equation}
\sum\limits_{a,b=0}^{1}(-1)^{ab}\langle m'n'\vert\hat{d}_{ab}(t)\vert mn
\rangle.
\end{equation}

\begin{figure}[h]
\includegraphics[width=7.4cm]{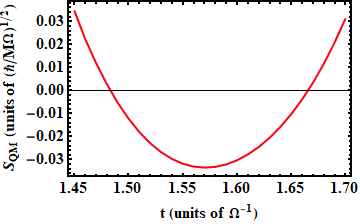}
\caption{Time-dependence of the Bell's parameter: the characteristic time scale of the system is set by the frequency $\Omega$. We are looking for the maximal violation at $T=\pi/2$ in units of $\Omega^{-1}$. Alice and Bob adopt the following strategy: for inputs $a,b=0$ they set their frequencies to $4\Omega$, whereas for $a,b=1$ they set them to $\Omega$. The endpoints of the interval during which the violation persists, i.e., $\mathcal{S}_{QM}(t)<0$, are $t_{i}\approx 1.485$ and $t_{f}\approx 1.665$. The maximal violation of the classical bound (which is zero), i.e., the minimal negative eigenvalue of $\hat{\mathcal{S}}_{QM}(T)$, is approximately $-0.034\times(\hbar/M\Omega)^{1/2}$.}
\label{results1}
\end{figure}

One only has to find a particular instant of time $t=T$ for which there exists an eigenvalue of $\hat{\mathcal{S}}_{QM}(T)$ that is smaller than zero (classical bound). The eigenstate of $\hat{\mathcal{S}}_{QM}(T)$ for this particular eigenvalue will be our initial state $\vert\Psi_{0}\rangle$. From the continuity of $\mathcal{S}_{QM}(t)$ follows that $S_{QM}(t)<0$ also in some interval $[t_{i},t_{f}]$ around $t=T$, implying the violation of the Bell's inequality for trajectories at least during that particular interval of time.  

To illustrate the above-described procedure, we provide, in Fig. \ref{results1}, a graphical representation of the evolution of Bell's parameter $\mathcal{S}_{QM}(t)$ in time. The violation in this case is not particularly strong, but it is a proof of principle that quantum mechanics allows such a violation. Details of the calculation are given in Sec. IV of our Supplemental Material. It would be desirable for future work to find further physical examples that lead to stronger violations, perhaps by choosing a different ansatz for the initial state or considering some other functional of the trajectories. 

\section{Conclusions and outlook}

In this paper, we have derived a \emph{Bell's theorem for trajectories}. Were trajectories fully predetermined--as assumed in classical physics--their pieces would clearly be predetermined, too. We have shown that quantum mechanics precludes this possibility, at least for some finite, continuous interval of time during the evolution.
As a matter of fact, quantum mechanics (by means of Bell's theorem) gave us good reasons to question the existence of predetermined values of physical observables. We showed that this argument has even more severe consequences, for it can be extended to \emph{prima facie} unobservable quantities: trajectories of particles in general do not exist predetermined.\\[0.2cm]

\section{Acknowledgements} 
The authors thank \v{C}aslav Brukner for fruitful discussions. 
D.G. and A.D. acknowledge support from the bilateral project SRB 02/2018 between Austria and Serbia. 
D.G. acknowledges support from the project no. ON171031 of Serbian Ministry of Education and Science. A.D. acknowledges
support from the project no. ON171035 of Serbian Ministry of Education and Science and from the scholarship
awarded by The Austrian Agency for International Cooperation in Education and Research (OeAD-GmbH).
F.D.S. acknowledges financial support
through a DOC Fellowship of the Austrian Academy of
Sciences (\"{O}AW). B.D. acknowledges support from an ESQ Discovery
Grant of the Austrian Academy of Sciences (\"{O}AW)
and the Austrian Science Fund (FWF) through BeyondC (F71).

\maketitle
\numberwithin{equation}{section}
\numberwithin{figure}{section}

\onecolumngrid
\setcounter{section}{0}

\section{Hidden variable model of single particle dynamics}
Here we present a general argument for the existence of a \emph{hidden variable} (HV) model of single-particle dynamics in terms of trajectories. Generally, in hidden variable theories, the quantum state is represented as an ensemble of classical (true) states in terms of hidden variables 
whose values determine, with certainty, the outcomes of individual measurements. The dispersion of measurement results occurs solely due to our ignorance, which is reflected through some (purely epistemological) probability distribution over the possible values of the hidden variables. \\

Let's suppose, for simplicity, that we have a single point-like particle moving on a discrete one-dimensional lattice of possible particle's positions $m\in\{1,\hdots,M\}$ each with spacing $\Delta x$. The length of the lattice is $L=(M-1)\Delta x$ (Fig. I.$1$). Initially, at $t_{0}=0$, the particle is in a position eigenstate $\vert m_{0}\rangle$ that subsequently evolves according to some arbitrary evolution $\hat{U}(t)$, during a time interval $t\in[0,T]$. 
We perform a sequence of $N$ projective measurements of particle's position at successive time instants $0< t_{1}< t_{2}<\hdots< t_{N}=T$ (not necessarily equidistant) to get the quantum mechanical probability distribution $P_{QM}(m_{1},\hdots,m_{N}\vert m_{0})$ of a sequence $(m_{1},\hdots,m_{N})$ of particle's positions, i.e., a ``discrete'' trajectory.
\begin{figure}[h!]
\begin{center}
\includegraphics[scale=0.45]{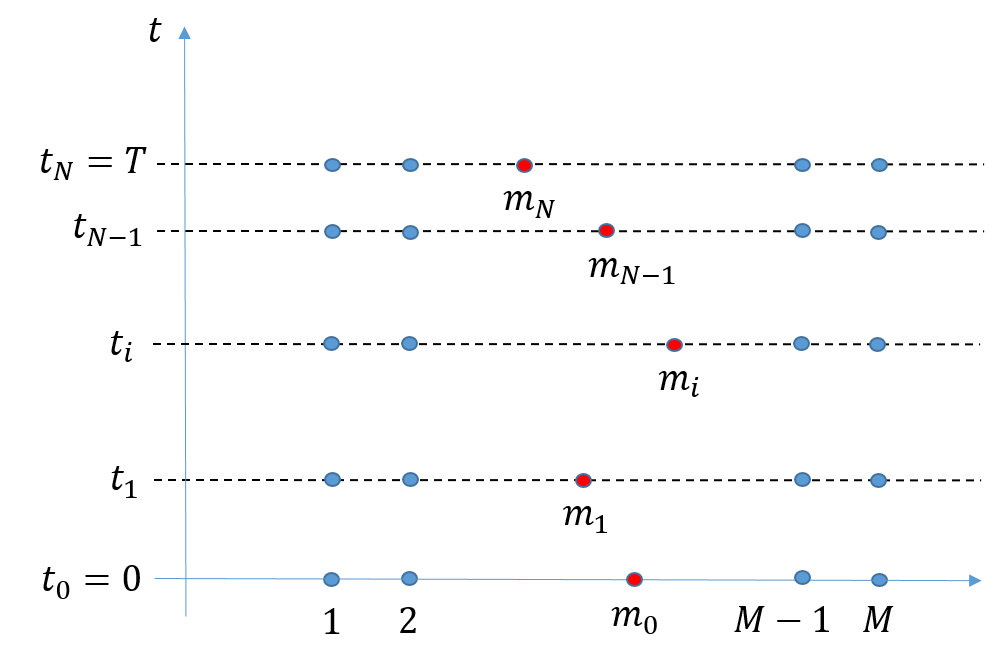}
\caption{Particle's evolution on a discrete one dimensional lattice: there are $M$ possible equidistant positions, $1$ to $M$, that particle can assume. Particle starts in some position eigenstate $\vert m_{0}\rangle$. According to quantum mechanics, between every two position measurements, performed at $t_{i-1}$ and $t_{i}$ ($i=1,\hdots,N$), particle's state evolves into a quantum superposition of position eigenstates, and during that time particle's position is undetermined. The result of the experiment is a particular sequence $(m_{1},\hdots,m_{N})$ of measurement results.}
\end{center}\label{Grid}
\end{figure}

Standard quantum mechanics gives us 
\begin{equation}
P_{QM}(m_{1},\hdots,m_{N}\vert m_{0})=\Vert \hat{P}_{m_{N}}\hat{U}_{N,N-1}\hdots\hat{P}_{m_{2}}\hat{U}_{2,1}\hat{P}_{m_{1}}\hat{U}_{1,0}\vert m_{0}\rangle\Vert^{2},
\end{equation}
where we have defined the evolution operators $\hat{U}_{i,i-1}\equiv\hat{U}(t_{i}-t_{i-1})$ and the projectors $\hat{P}_{m_{i}}=\vert m_{i}\rangle\langle m_{i}\vert$ ($i=1,\hdots,N$) on position eigenstates $\vert m_{i}\rangle$. In terms of propagators (transition amplitudes) $K_{m_{i},m_{i-1}}\equiv\langle m_{i}\vert \hat{U}_{i,i-1}\vert m_{i-1}\rangle$ the quantum probability reads
\begin{equation}\label{PQM}
P_{QM}(m_{1},\hdots,m_{N}\vert m_{0})=\prod\limits_{i=1}^{N}\vert K_{m_{i},m_{i-1}}\vert^{2}.
\end{equation} 
Obviously, $\sum\limits_{m_{1}=1}^{M}\hdots\sum\limits_{m_{N}=1}^{M}P_{QM}(m_{1},\hdots,m_{N}\vert m_{0})=1$.
Equation (\ref{PQM}) already suggests that the distribution of measurement results can be seen as a classical Markov chain. Our goal is to show that this distribution indeed admits a deterministic (classical) model.\\

In general, each $M\times M$ matrix $\mathbb{K}^{(i)}$ ($i=1,\hdots,N$) whose entries $[\mathbb{K}^{(i)}]_{j_{i},j_{i-1}}$ are transition probabilities $\vert K_{j_{i},j_{i-1}}\vert^{2}=\vert\langle j_{i}\vert \hat{U}_{i,i-1}\vert j_{i-1}\rangle\vert^{2}$, with $j_{i-1},j_{i}\in\{1,\hdots,M\}$, is a \emph{doubly stochastic matrix}, i.e., a matrix with real, non-negative entries whose each column and row adds up to one. According to the \emph{Birkhoff-von Neumann theorem}, every matrix of this kind can be represented as a convex sum of $M\times M$ permutation matrices. In other words, the class of $M\times M$ doubly stochastic matrices forms a \emph{convex polytope} (known as the Birkhoff polytope) whose vertices are the permutation matrices, in particular
\begin{equation}
\mathbb{K}^{(i)}=\sum\limits_{\pi_{i}\in S_{M}}\mu^{(i)}_{\pi_{i}}\mathcal{P}_{\pi_{i}},
\end{equation}
where $S_{M}=\{\pi^{(a)}:\{1,\hdots,M\}\rightarrow\{1,\hdots,M\}
; \;a=1,\hdots,M!\}$ is the group of permutations of $M$ elements and $\mathcal{P}_{\pi_{i}}$ is an $M\times M$ permutation matrix composed of zeros and ones, that corresponds to the permutation $\pi_{i}$ that acts at $i^{th}$ step, i.e., at $t=t_{i-1}$. Since $\mathbb{K}^{(i)}$ is a convex sum, we have $\sum_{\pi_{i}}\mu^{(i)}_{\pi_{i}}=1$ and 
$\mu^{(i)}_{\pi_{i}}\geq 0$ for all $\pi_{i}$. \\

At $t=t_{i-1}$ particle is found at $m_{i-1}$ and its subsequent evolution is determined by $\mathbb{K}^{(i)}$. This matrix gives us probabilities to go from $m_{i-1}$ at $t_{i-1}$ to other positions at $t_{i}$. Each permutation matrix $\mathcal{P}_{\pi_{i}}$ corresponds to a deterministic transition, and the $\mathbb{K}^{(i)}$ matrix can be regarded as a convex mixture of deterministic processes the probability of each being the corresponding coefficient $\mu^{(i)}_{\pi_{i}}$ in the convex sum. In particular, the probability of $m_{i-1}\rightarrow m_{i}$ transition is
\begin{equation}
\vert K_{m_{i},m_{i-1}}\vert^{2}=\sum\limits_{\pi_{i}\in S_{M}}\mu^{(i)}_{\pi_{i}}[\mathcal{P}_{\pi_{i}}]_{m_{i},m_{i-1}}=\sum\limits_{\pi_{i}\in S_{M}}\mu^{(i)}_{\pi_{i}}\delta_{m_{i},\pi_{i}(m_{i-1})}=
\sum\limits_{\{\pi_{i}\in S_{M}\;\vert\;\pi_{i}(m_{i-1})=m_{i}\}}\mu^{(i)}_{\pi_{i}},
\end{equation}
i.e., we have to take into account every permutation that maps $m_{i-1}$ to $m_{i}$, and there are $(M-1)!$ such permutations for each transition. 
The joint probability now reads
\begin{align}\label{PQM_HV}
P_{QM}(m_{1},\hdots,m_{N}\vert m_{0})=
\prod\limits_{i=1}^{N}\vert K_{m_{i},m_{i-1}}\vert^{2}=\prod\limits_{i=1}^{N}\sum\limits_{\pi_{i}\in S_{M}}\mu^{(i)}_{\pi_{i}}\delta_{m_{i},\pi_{i}(m_{i-1})}.
\end{align} 
Each sequence of permutations $(\pi_{1},\hdots,\pi_{N})$ corresponds to a particular deterministic evolution $m_{0}\rightarrow\pi_{1}(m_{0})\rightarrow\pi_{2}(\pi_{1}(m_{0}))\rightarrow\hdots\rightarrow \pi_{N}(\pi_{N-1}(\hdots\pi_{2}(\pi_{1}(m_{0}))))$ with the probability given by the product $\mu^{(1)}_{\pi_{1}}\hdots\mu^{(N)}_{\pi_{N}}$. 
These permutations themselves can be seen as ``hidden variables'' that determine completely particle's trajectory.\\

By taking $M,N\rightarrow \infty$, while keeping $L$ and $T$ fixed, we obtain a continuous limit, and the quantum mechanical probability density to observe a particular sequence $(x_{1},\hdots,x_{N})$ of particle's positions can be represented, for example, by double path integral \cite{Dowker}. 
In the HV model, particle has a definite position at every $t\in[0,T]$, i.e., a trajectory. When we perform a position measurement we just ``reveal'' the particle's position at that particular time instant. Which way the particle goes next depends on the value of the hidden variable (Fig. I.$2$). While the concrete sequence of position-measurement outcomes is completely determined by the values of the HV, their probabilistic nature (in the repeated experiment) appears because of the probability distribution over the possible values of the HV (which is entirely due to our ignorance).

\begin{figure}[h!]
\begin{center}
\includegraphics[scale=0.38]{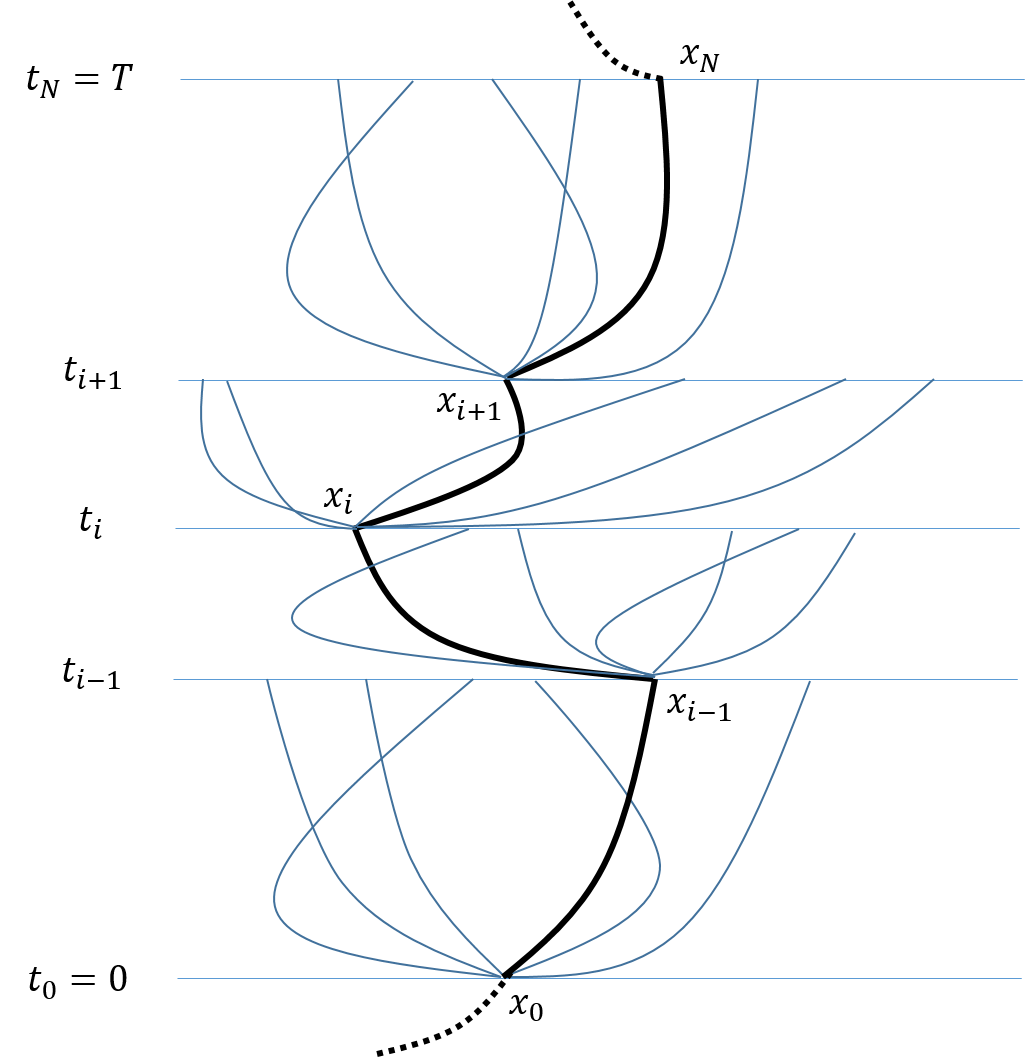}
\caption{Continuous version of a single-particle dynamics in HV interpretation: particle follows an underlying trajectory (black line) during the time interval $[0,T]$. After each position measurement, the HV tells the particle where to go next. However, there is a probability distribution over the set of all possible values of the HV which induces a probability distribution over the subsequent paths (blue lines)--other ``permutations''--that particle could take between two position measurements.}
\end{center}
\end{figure}

\section{Proof of the general Bell's inequality}
We prove that inequality $(5)$ holds for any symmetric, subadditive functional, i.e., $F[X]=F[-X]$ and $F[X+Y]\leq F[X]+F[Y]$. Let us start by explicitly writing the averages appearing in the inequality $(5)$, by plugging Eq. $(4)$ in it:
\begin{equation}
S=\int d\lambda_Ad\lambda_B\;\mu(\lambda_A,\lambda_B)\sum_{a,b=0}^1(-1)^{ab}F[X^{(a)}_{\lambda_A}-Y^{(b)}_{\lambda_B}].
\end{equation}
To prove that $S\geq 0$, it is sufficient to demonstrate that the sum over $a,b$ is non-negative (since $\mu\geq 0$, being a distribution), i.e.,
\begin{equation}\label{bellproof}
\begin{aligned}
\sum_{a,b=0}^1(-1)^{ab}F[X^{(a)}_{\lambda_A}-Y^{(b)}_{\lambda_B}]&=F[X^{(0)}_{\lambda_A}-Y^{(0)}_{\lambda_B}]
+F[X^{(0)}_{\lambda_A}-Y^{(1)}_{\lambda_B}]
+F[X^{(1)}_{\lambda_A}-Y^{(0)}_{\lambda_B}]
-F[X^{(1)}_{\lambda_A}-Y^{(1)}_{\lambda_B}]\geq0.
\end{aligned}
\end{equation}
In order to show this, let us relabel the arguments of the functional in the previous expression as $Z_{00}$, $Z_{01}$, $Z_{10}$ and $Z_{11}$, respectively, from the left to the right. (For example, $Z_{00}:=X^{(0)}_{\lambda_A}-Y^{(0)}_{\lambda_B}$). In this way, the inequality \eqref{bellproof} reads
\begin{equation}
F[Z_{11}]\leq F[Z_{00}]+F[Z_{01}]+F[Z_{10}].
\end{equation}
The symmetry of $F$ ensures $F[Z_{00}]=F[-Z_{00}]$, which together with the  subadditivity completes the proof:
\begin{equation}
F[Z_{11}]=F[-Z_{00}+Z_{01}+Z_{10}]\leq F[-Z_{00}]+F[Z_{01}]+F[Z_{10}]=F[Z_{00}]+F[Z_{01}]+F[Z_{10}].
\end{equation}
\section{Separable states and the Bell inequality for trajectories}
Here we give a short proof that separable states, in general, cannot bring about a violation of the Bell's inequality for trajectories. 
Let us suppose that a pair of particles (denoted ``A'' and ``B'') is initially (at $t=0$) prepared in a convex mixture of separable states, 
\begin{equation}
\hat{\rho}_{AB}(t=0)=\sum\limits_{s}\mu_{s}\;\hat{\rho}^{(s)}_{A}\otimes\hat{\rho}^{(s)}_{B},
\end{equation} 
with $\sum\limits_{s}\mu_{s}=1$ and all $\mu_{s}\geq 0$.

Take a pair of general POVM operators (to take into account  imperfect measurement; ideally, we would use sharp projectors) $\hat{M}^{(a)}_{A}(x,t)$ and $\hat{N}^{(b)}_{B}(x,t)$ that act on particles A and B, respectively, satisfying the normalization conditions:
\begin{equation}
\int\limits_{-\infty}^{+\infty}dx\;\hat{M}^{(a)}_{A}(x,t)=\hat{\mathbb{I}}_{A},\;\;\;\;\;\;
\int\limits_{-\infty}^{+\infty}dy\;\hat{N}^{(b)}_{B}(y,t)=\hat{\mathbb{I}}_{B}.
\end{equation}  
They depend on inputs $a,b\in\{0,1\}$ and evolve, in Heisenberg picture, as
\begin{equation}
\hat{M}^{(a)}_{A}(x,t)=\hat{L}^{(a)\dagger}_{A}(t)\hat{M}_{A}\hat{L}^{(a)}_{A}(t),\;\;\;\;\;\;\hat{N}^{(b)}_{B}(y,t)=\hat{L}^{(b)\dagger}_{B}(t)\hat{N}_{B}\hat{L}^{(b)}_{B}(t),
\end{equation}
where we introduced linear evolution operators $\hat{L}^{(a)}_{A}(t)$ and $\hat{L}^{(b)}_{B}(t)$ (to take into account possible dissipation/decoherence effects; ideally, they would be unitary).  

Conditional probability for a pair of measurement outcomes $(x_{a},y_{b})$, given the inputs $(a,b)$, is
\begin{equation}
p(x_{a},y_{b}\vert a,b)=Tr\left(\hat{\rho}_{AB}(0)\hat{M}^{(a)}_{A}(x_{a},t)\otimes\hat{N}^{(b)}_{B}(y_{b},t)\right).
\end{equation}
Now, in the spirit of Fine's theorem \cite{Fine}, we introduce a joint probability distribution:
\begin{equation}
p(x_{0},x_{1},y_{0},y_{1})=\sum\limits_{s}\mu_{s}\;p^{(s)}_{A}(x_{0}\vert 0)p^{(s)}_{A}(x_{1}\vert 1)p^{(s)}_{B}(y_{0}\vert 0)p^{(s)}_{B}(y_{1}\vert 1), 
\end{equation} 
which clearly gives us the appropriate marginal probabilities, e.g., $\int dx_{1}\int dy_{1}\;p(x_{0},x_{1},y_{0},y_{1})=p(x_{0},y_{0}\vert 0,0)$. Therefore, we see that separable states admit local probability decomposition.

Quantum Bell's parameter at a particular time instant $t$ is given by
\begin{align}
\mathcal{S}_{QM}(t)&=\sum\limits_{a,b=0}^{1}(-1)^{ab}\;Tr(\hat{\rho}_{AB}(t)\hat{f}_{ab}(\hat{x}-\hat{y}))\nonumber\\
&=\int\limits_{-\infty}^{+\infty}dx_{0}\int\limits_{-\infty}^{+\infty}dx_{1}\int\limits_{-\infty}^{+\infty}dy_{0}\int\limits_{-\infty}^{+\infty}dy_{1}\;p(x_{0},x_{1},y_{0},y_{1})[f(x_{0}-y_{0})+f(x_{0}-y_{1})+f(x_{1}-y_{0})-f(x_{1}-y_{1})]. 
\end{align}
Finally, the term in the square brackets is greater or equal to zero, which is a direct consequence of the triangle inequality, and so the same holds for the whole quantum Bell's parameter at every instant of time. This means that Bell's inequality for trajectories cannot be violated using separable states.   

\section{Procedure of finding the initial state}
Here we present, in some more detail, the procedure of finding the initial state $\vert\Psi_{0}\rangle$ that ensures a violation of the inequality $\mathcal{S}(t)>0$ at a particular instant of time $t=T$, which immediately extends to some finite continuous interval $[t_{i},t_{f}]$ such that $T\in[t_{i},t_{f}]$, due to continuity of $\mathcal{S}(t)$. For that, we make a transition to the Heisenberg picture.

Under the action of one-dimensional harmonic oscillator Hamiltonian, with some generic frequency $\omega$, the coordinate and momentum operators for Alice's particle, $\hat{x}(t)$ and $\hat{p}_{x}(t)$ (and likewise $\hat{y}(t)$ and $\hat{p}_{y}(t)$ for Bob's particle), evolve in time according to the Heisenberg's equations:
\begin{eqnarray}
\hat{x}(t)&=&\hat{x}(0) \cos (\omega t) + \frac{\hat{p}_{x}(0)}{M\omega}\sin(\omega t) ,\\
\hat{p}_{x}(t)&=&\hat{p}_{x}(0) \cos (\omega t) -M\omega \hat{x}(0)\sin(\omega t).
\end{eqnarray}
For $\omega t=2\pi$, this ``rotation'' reduces to the identity transformation, whereas for $\omega t=\pi/2$ we get the interchange of the operators -- the coordinate operator becomes $\hat{p}(0)/M\omega$ and the momentum operator becomes $-M\omega\hat{x}(0)$ -- because the transformation is just an ordinary Fourier transform,
\begin{equation}\label{F}
\hat{F}_{A/B}=\mathrm{e}^{-i\frac{\pi}{2}(\hat{N}_{A/B}+\frac{1}{2})},
\end{equation}
where $\hat{N}_{A/B}$ is the number operator. 

The evolution of the operator $\hat{d}(\hat{x}-\hat{y})$ for a given pair of inputs $(a,b)$ is 
\begin{equation}
\hat{d}_{ab}(t)=\mathrm{e}^{\frac{i}{\hbar} \hat{H}_{B}^{(b)} t} \mathrm{e}^{\frac{i}{\hbar} \hat{H}_{A}^{(a)} t} \hat{d}(\hat{x}-\hat{y}) \mathrm{e}^{-\frac{i}{\hbar} \hat{H}_{A}^{(a)} t} \mathrm{e}^{-\frac{i}{\hbar} \hat{H}_{B}^{(b)} t},
\end{equation}
with $\hat{H}_{A/B}^{(a/b)}=\hbar\omega_{a/b}(\hat{N}^{(a/b)}+1/2)$.

For a given frequency $\Omega$, let us fix a particular instant of time, say $t=T=(\pi/2)\Omega^{-1}$, at which we want to obtain the maximal violation. Alice and Bob agree to set their frequencies to $\Omega$ for the input value $1$, and to set them to $4\Omega$ if for the input value $0$. This leads to four possible cases:
\begin{enumerate}
    \item For $(a,b)=(0,0)$ we have $(\omega_{a},\omega_{b})=(4\Omega,4\Omega)$ and hence $\omega_{a}T=\omega_{b}T=2\pi$. Therefore,
    \begin{equation}
    \hat{d}_{00}(T)=\hat{d}(\hat{x}-\hat{y}).
    \end{equation}
   
    \item For $(a,b)=(0,1)$ we have $(\omega_{a},\omega_{b})=(4\Omega,\Omega)$ and hence $\omega_{a}T=2\pi$ and $\omega_{b}T=\pi/2$. Therefore,
    \begin{equation}
    \hat{d}_{01}(T)=\hat{F}^{\dag}_{B}\hat{d}(\hat{x}-\hat{y})\hat{F}_{B}.
    \end{equation}
   
    \item For $(a,b)=(1,0)$ we have $(\omega_{a},\omega_{b})=(\Omega,4\Omega)$ and hence $\omega_{a}T=\pi/2$ and $\omega_{b}T=2\pi$. Therefore,
    \begin{equation}
    \hat{d}_{10}(T)=\hat{F}^{\dag}_{A}\hat{d}(\hat{x}-\hat{y})\hat{F}_{A}.
    \end{equation}
    
   \item For $(a,b)=(1,1)$ we have $(\omega_{a},\omega_{b})=(\Omega,\Omega)$ and hence $\omega_{a}T=\omega_{b}T=\pi/2$. Therefore, 
    \begin{equation}
    \hat{d}_{11}(T)=\hat{F}^{\dag}_{B}\hat{F}^{\dag}_{A}\hat{d}(\hat{x}-\hat{y})\hat{F}_{A}\hat{F}_{B}.
    \end{equation}
We assume that the initial state $\vert\Psi_{0}\rangle$ belongs to the subspace spanned by $\{\vert m\rangle_{A}\vert n\rangle_{B} \;\vert\; m,n=0,1,\dots,8\}$, where $\vert m\rangle_{A}$ and $\vert n\rangle_{B}$ are energy eigenstates for frequency $\Omega$. In this basis, operators $\hat{F}_{A}$ and $\hat{F}_{B}$ are represented by the following matrices:
\begin{equation}\label{FA,FB}
F_{A}= 
  \begin{bmatrix}
    1 & & \\
    & -i &    \\
     & & \ddots & \\
    & & & (-i)^{8}
  \end{bmatrix}
\otimes\mathbb{I}_{9\times 9} ,\;\;\;F_{B}=\mathbb{I}_{9\times 9}\otimes  \begin{bmatrix}
    1 & & \\
    & -i &    \\
     & & \ddots & \\
    & & & (-i)^{8}
  \end{bmatrix}.
\end{equation}
Finally, the Bell's operator at $t=T$ is \begin{align}
\hat{\mathcal{S}}_{QM}(T)=\sum\limits_{a,b=0}^{1}(-1)^{ab}\hat{d}_{ab}(T) =\hat{d}(\hat{x}-\hat{y})+\hat{F}^{\dag}_{B}\hat{d}(\hat{x}-\hat{y})\hat{F}_{B}+\hat{F}^{\dag}_{A}\hat{d}(\hat{x}-\hat{y})\hat{F}_{A}-\hat{F}^{\dag}_{B}\hat{F}^{\dag}_{A}\hat{d}(\hat{x}-\hat{y})\hat{F}_{A}\hat{F}_{B},
\end{align}
and it is represented by a certain $81\times 81$ matrix. By solving the eigenvalue problem of this matrix (using \textit{Wolfram Mathematica}) we find that its spectrum has a minimal negative eigenvalue $\xi_{-}\approx-0.034$, the unit of length being $(\hbar/M\Omega)^{1/2}$, which depends on the parameters $M$ and $\Omega$. 

If we now identify the initial state $\vert\Psi_{0}\rangle$ with the (entangled) eigenstate $\vert\xi_{-}\rangle$, our procedure ensures a violation at $t=T$, i.e., 
\begin{equation}
\mathcal{S}_{QM}(T)=\sum\limits_{a,b=0}^{1}(-1)^{ab}\langle\Psi_{AB}^{(a,b)}(T)\vert\hat{d}(\hat{x}-\hat{y})\vert\Psi_{AB}^{(a,b)}(T)\rangle<0.   
\end{equation}

Since $\mathcal{S}_{QM}(t)$ is a continuous function of time, we expect also to have $\mathcal{S}_{QM}(t)<0$ in some finite continuous interval $[t_{i},t_{f}]$ such that $T\in[t_{i},t_{f}]$, hence $S=\int_{t_{i}}^{t_{f}}dt\mathcal{S}_{QM}(t)<0$. Having selected the appropriate initial state, we can propagate it from $t=0$ to any $t>0$, for all four instances of $(a,b)$, and calculate analytically the whole function $\mathcal{S}_{QM}(t)$; in particular, we can find $t_{i}$ and $t_{f}$. We conclude that the pieces of the trajectories of the particle's that correspond to the interval $[t_{i},t_{f}]$ cannot be accounted for by any theory that assumes local realism. In our case, $T=\tfrac{\pi}{2}\;\Omega^{-1}$, $t_{i}\approx 1.485 \;\Omega^{-1}$ and $t_{f}\approx 1.665\;\Omega^{-1}$, see Fig. $3$. 

\end{enumerate}

\end{document}